# PROBABILISTIC PROJECTIONS OF HIV PREVALENCE USING BAYESIAN MELDING[1]

By Leontine Alkema, Adrian E. Raftery and Samuel J. Clark

*University of Washington*

The Joint United Nations Programme on HIV/AIDS (UNAIDS) has developed the Estimation and Projection Package (EPP) for making national estimates and short-term projections of HIV prevalence based on observed prevalence trends at antenatal clinics. Assessing the uncertainty about its estimates and projections is important for informed policy decision making, and we propose the use of Bayesian melding for this purpose. Prevalence data and other information about the EPP model's input parameters are used to derive a probabilistic HIV prevalence projection, namely a probability distribution over a set of future prevalence trajectories. We relate antenatal clinic prevalence to population prevalence and account for variability between clinics using a random effects model. Predictive intervals for clinic prevalence are derived for checking the model. We discuss predictions given by the EPP model and the results of the Bayesian melding procedure for Uganda, where prevalence peaked at around 28% in 1990; the 95% prediction interval for 2010 ranges from 2% to 7%.

**1. Introduction.** In this article we propose a way to obtain probabilistic projections of HIV prevalence for generalized epidemics in countries with little detailed knowledge of HIV prevalence. A *generalized* HIV epidemic affects the general population beyond high risk sub-populations. To qualify as generalized under WHO and UNAIDS definitions, an epidemic must affect one percent or more of pregnant women [Ghys et al. (2004)] and, by this definition, many countries in sub-Saharan Africa have generalized HIV/AIDS epidemics. In countries with high HIV prevalence the effects of the epidemic

Received February 2007; revised March 2007.
[1]Supported by a seed grant from the Center for Statistics and the Social Sciences (CSSS), the Center for Studies in Demography and Ecology (CSDE) and the Blumstein-Jordan Professorship.

on the population are significant, and prevalence predictions are necessary to understand and plan for these effects in the future.

UNAIDS has to produce current estimates and short-term projections of HIV prevalence for all countries with generalized HIV epidemics. It developed the Estimation and Projection Package (EPP) for this purpose. EPP is designed to reproduce the overall dynamics of a generalized HIV epidemic in a parsimonious way. To achieve this, the EPP model must be general, robust and simple; simple because many countries with generalized HIV epidemics have few data to describe those epidemics. For most countries in sub-Saharan Africa, the main source of information on HIV prevalence is the prevalence of HIV among women who attend antenatal clinics. As a result, the EPP model is designed to produce past and future trends in HIV prevalence that are consistent with measured trends in antenatal clinic prevalence. For a small number of countries, more data of higher quality are available. It is conceivable that for those countries one could design a model that would be better able to capture the complex dynamics revealed by the data. However, this would violate the UNAIDS requirement that a single consistent method be used to produce estimates and projections of HIV prevalence for all affected countries, and could confuse comparison of the estimates between countries, and across time for the same country.

The EPP 2005 software and supporting documentation can be downloaded from www.unaids.org. EPP is used by UNAIDS, the World Health Organization and a variety of national and other agencies to produce HIV prevalence estimates. EPP is often used in combination with Spectrum, an extended sex- and age-differentiated population model [Stover (2004), Stover et al. (2006)]. Spectrum uses the prevalence trends generated by EPP to produce annual sex- and age-specific HIV incidence, deaths and other quantities of interest. EPP and Spectrum together have been used to analyze the global impact of HIV/AIDS prevention and treatment programs, and the World Health Organization has proposed using their combined output to estimate the number of people in need of antiretroviral treatment.

Uncertainty is inherent in forecasts of future trends in HIV prevalence, and assessing it is crucial for policy decision making. In 2003, UNAIDS included information on uncertainty in its estimates and projections by calculating and presenting plausibility bounds [Grassly et al. (2004)]. These bounds were derived by combining the results of a bootstrap method with expert opinion regarding the range of possible epidemic curves. As noted by Morgan et al. [(2006), page iii77]: "plausibility bounds do not represent and should not be interpreted as formal statistical confidence intervals." A Bayesian framework solves this problem by providing a way of including expert opinion while still giving formal statistical confidence intervals.

We propose Bayesian melding to obtain probabilistic projections of HIV prevalence. Bayesian melding was first developed to estimate the rate of increase of whale populations [Raftery et al. (1995), Poole and Raftery (2000)], and was successfully applied to policy-making in that context. Bayesian melding fully accounts for information describing uncertainty in both the inputs and outputs of a deterministic model. In this paper we discuss the application of the Bayesian melding procedure to the EPP model. In Section 2 we describe the EPP model, in Section 3 we explain the Bayesian melding procedure and a random effects model for HIV prevalence, in Section 4 we present results for urban HIV prevalence in Uganda, and in Section 5 we discuss possible improvements to the methodology.

**2. The Estimation and Projection Package.** The Estimation and Projection Package (EPP) was developed by UNAIDS to satisfy two major constraints. First it has to be able to capture the main dynamics of an HIV epidemic in any country, often without detailed knowledge of the transmission patterns in that country, and second it has to be simple enough to be used by national planning officials in a wide variety of developing countries. The UNAIDS Reference Group on Estimates, Modeling and Projections (2002) recognized that there are many complex models that incorporate patterns of risk behavior and mixing, and provide useful tools for understanding the spread and control of HIV. Many of these models are unsuitable for the task at hand, however, because they require a large number of biological and behavioral parameter values that are not available from all the countries that require estimates and projections of HIV prevalence.

UNAIDS has developed a simple epidemiological Susceptible-Infected model that satisfies these two constraints. The population at time $t$ is divided into three groups, a not-at-risk group $X(t)$, an at-risk group $Z(t)$ and an infected group $Y(t)$. The model assumes a constant non-AIDS mortality rate $\mu$ and fertility rate $b$ and does not represent migration or age structure. Time-evolving prevalence (the fraction of the population infected) is modeled with four parameters that need to be estimated, $r$, $f_0$, $t_0$ and $\phi$. The parameter $r$ is the rate of infection, $f_0$ is the fraction of the population in the at-risk category at the start of the epidemic, $t_0$ is the start time of the epidemic and $\phi$ is the behavioral response.

The influence of each of these parameters on the shape of the epidemic is shown in Figure 1. Note the standard overall shape of the epidemic in all of the plots. The fraction of the HIV negative population infected each year, or incidence, increases to a maximum and declines thereafter. As long as the number of new HIV infections is greater than the number of AIDS deaths, the prevalence rate increases. An epidemic peaks when the incidence and mortality rates are about equal. After the peak the prevalence rate comes down because the number of AIDS deaths continues to increase as a result of the lag between becoming infected and dying of AIDS. An epidemic stabilizes when the number of new HIV infections equals the number of AIDS deaths.

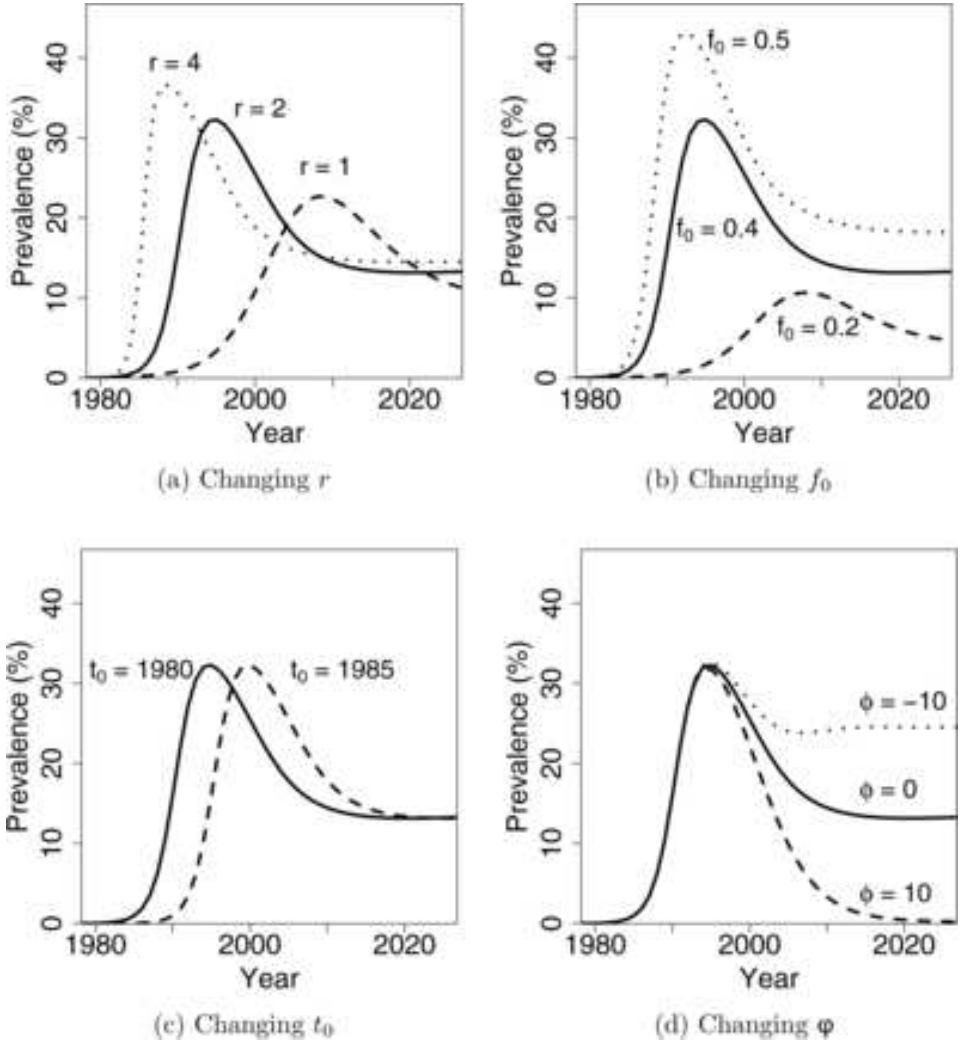

Fig. 1. *The influence of $r$, $f_0$, $t_0$ or $\phi$ on the shape of the epidemic curve while holding other parameters fixed at the values $r = 2$, $f_0 = 0.4$, $t_0 = 1980$, $\phi = 0$.*

The greater the rate of infection $r$, the faster prevalence increases at the beginning of the epidemic, as can be seen from Figure 1(a). If a fraction $f$ of the population is at risk of HIV infection, every infected person infects $r \cdot f$ people each year ($r \cdot f$ is called the force of infection). The fraction of the population in the at-risk category at the start of the epidemic $f_0$ influences when and at what prevalence the epidemic peaks. If a higher proportion of the population is initially at risk, the epidemic will peak sooner and at a higher level, as can be seen in Figure 1(b). The epidemic is shown for two different start years in Figure 1(c). Changing the start year, $t_0$, does not change the shape of the epidemic, only its timing.

The behavioral response, $\phi$, influences the level at which the epidemic levels off after the peak, as shown by Figure 1(d). The parameter $\phi$ is positively related to steady state prevalence. Negative values of $\phi$ correspond to a situation in which new members of the population change their behavior when they see others dying of AIDS, so that fewer of them enter the at-risk group. Positive values of $\phi$ correspond to the opposite situation in which a larger fraction of new members enter the at-risk group. This could be due to pockets of the population that were previously isolated, perhaps by geography or culture, being exposed to infection.

The rates at which the sizes of the groups change—through recruitment, behavior and infection—are described by three differential equations:

$$\begin{cases} \dfrac{dX(t)}{dt} = \left(1 - f\left(\dfrac{X(t)}{N(t)}, f_0, \phi\right)\right) E(t) - \mu X(t), \\ \dfrac{dZ(t)}{dt} = f\left(\dfrac{X(t)}{N(t)}, f_0, \phi\right) E(t) - \left(\mu + r\dfrac{Y(t)}{N(t)} + \lambda(t)\right) Z(t), \\ \dfrac{dY(t)}{dt} = \left(r\dfrac{Y(t)}{N(t)} + \lambda(t)\right) Z(t) - \int_0^t \left(r\dfrac{Y(\tau)}{N(\tau)} + \lambda(\tau)\right) Z(\tau) g(t - \tau) \, d\tau, \end{cases}$$

where $N(t)$ is the total population, $N(t) = X(t) + Z(t) + Y(t)$, and $\mu$ is the nonHIV death rate. The function $g(\tau)$ specifies the HIV death rate $\tau$ years after infection. Survival after infection is assumed to have a Weibull(2.4, 10.5) distribution, so that the median survival time is 9 years. The start year of the epidemic is defined as the first year in which people are infected; the time at which a fraction $\lambda_0$ of the at-risk group $Z$ moves to the infected group $Y$. Similar epidemics can be generated with either a larger, earlier pulse or a smaller, later pulse. We set $\lambda_0 = 0.001$ so that 0.1% of the at-risk population gets infected in the start year. The initial pulse is modeled such that $\lambda(t) = \lambda_0 \cdot \delta(t - t_0)$, where $\delta(t)$ is the Dirac Delta function, and so $\int_{-\infty}^{\infty} \lambda(t) Z(t) \, dt = \lambda_0 Z(t_0)$.

The population being modeled is aged 15+. New members of the population are the ones who survive to age 15. When entering the population, they are assigned to either the not-at-risk group $X(t)$ or the at-risk group $Z(t)$. The total number of new members at time $t$, $E(t)$, depends on the population size 15 years ago, the birth rate and the survival rate from birth to age 15. The birth rate is applied to both the uninfected and infected groups, taking into account the HIV-related fertility reduction experienced by the infected group and the transmission of HIV from mother to child. A fraction of the new 15-year-old members enter the at-risk group $Z(t)$ at time $t$. This fraction is given by $f(\frac{X(t)}{N(t)}, f_0, \phi)$ (defined later in Section 3.2), and depends on the fraction of the population in the not-at-risk group, the fraction initially at risk, and the behavioral response $\phi$. The remainder of the new 15-year-olds enter the not-at-risk group $X(t)$.

## 3. Bayesian melding for the EPP model.

3.1. *Bayesian melding.* The EPP model transforms the input $\boldsymbol{\theta}$, consisting of the four input parameters $(r, f_0, t_0, \phi)$, into the output $\boldsymbol{\rho}$, consisting of a series of HIV prevalence rates for the population during a given period. We denote the EPP model by $M$, so that $\boldsymbol{\rho} = M(\boldsymbol{\theta})$. The Bayesian melding procedure [Poole and Raftery (2000)] combines information on inputs and outputs. In the case of the EPP model, these consist of prior knowledge about the inputs, and prior knowledge and data informative about the outputs.

First consider the melding of all information about outputs. Let $\boldsymbol{W}$ denote the data that are informative about prevalence rates. The prevalence data yield a likelihood, $p(\boldsymbol{W}|\boldsymbol{\rho})$, for the model output $\boldsymbol{\rho}$. Expert knowledge provides a prior distribution $p(\boldsymbol{\theta})$ for the inputs. This prior density induces a prior density on the outputs $\boldsymbol{\rho}$, because $\boldsymbol{\rho}$ is a transformation of $\boldsymbol{\theta}$; we denote this *induced* prior on the outputs by $p^*(\boldsymbol{\rho})$.

Expert knowledge about prevalence is specified by a prior distribution of prevalence in certain years. Here this prior is taken to be independent between years, and uniform between upper and lower bounds for each year. For example, restricting prevalence in 1980 to be smaller than 10% is represented by the prior distribution $\rho_{1980} \sim U(0, 10)$. This results in a *direct* (as opposed to induced) prior distribution $p(\boldsymbol{\rho})$ on the outputs, given by

$$p(\boldsymbol{\rho}) \propto \prod_{t \in \Gamma} I_{V_t}(\rho_t),$$

where $\Gamma$ is the set of years for which there is prior knowledge, $V_t$ is the interval given by the lower and upper bounds on prevalence in year $t$, and $I_A(x)$ is the indicator function for a set $A$, equal to 1 if $x \in A$ and to 0 if not.

As proposed by Poole and Raftery (2000), the two prior distributions on the output, namely, the induced prior $p^*(\boldsymbol{\rho})$ and the direct prior $p(\boldsymbol{\rho})$, are combined using logarithmic pooling:

(1) $$\tilde{p}(\boldsymbol{\rho}) \propto p^*(\boldsymbol{\rho})^\alpha p(\boldsymbol{\rho})^{1-\alpha},$$

where $\tilde{p}(\boldsymbol{\rho})$ is the pooled prior and $\alpha$ is the pooling weight. Because the direct prior distributions are uniform on intervals, we let $\alpha \uparrow 1$; note that the resulting limiting pooled prior is not the same as that obtained by setting $\alpha = 1$ in (1). The resulting pooled prior on outputs incorporates the boundaries on outputs as given by expert opinion on outputs, while remaining maximally faithful to the prior induced by the inputs. The pooled prior of the output is then

$$\tilde{p}(\boldsymbol{\rho}) \propto p^*(\boldsymbol{\rho}) \prod_{t \in \Gamma} I_{V_t}(\rho_t).$$

The posterior distribution of the output is then given by

(2) $$p(\boldsymbol{\rho}|\boldsymbol{W}) \propto \tilde{p}(\boldsymbol{\rho}) p(\boldsymbol{W}|\boldsymbol{\rho}).$$

By similar reasoning, the posterior distribution of the inputs is given by

(3) $$p(\boldsymbol{\theta}|\boldsymbol{W}) \propto \tilde{p}(\boldsymbol{\theta}) p(\boldsymbol{W}|M(\boldsymbol{\theta})),$$

where $\tilde{p}(\boldsymbol{\theta})$ is the pooled prior on the inputs and $p(\boldsymbol{W}|M(\boldsymbol{\theta}))$ is the likelihood of the inputs. The pooled prior on the inputs, $\tilde{p}(\boldsymbol{\theta})$, is given by the limit of equation (16) of Poole and Raftery (2000) as $\alpha \uparrow 1$. In the present case, this simplifies to the direct prior density restricted to the region of input parameter space where the constraints on outputs are satisfied, and rescaled accordingly.

It is not possible to write the posterior distributions in (2) and (3) analytically because the model is not invertible. We approximate it using Monte Carlo methods, by drawing a random sample from it using the Sampling Importance Resampling algorithm [Rubin (1987, 1988)]:

1. Sample $\{\boldsymbol{\theta}^{(1)}, \ldots, \boldsymbol{\theta}^{(n)}\}$ from the input prior $p(\boldsymbol{\theta})$ on $\boldsymbol{\theta} = (r, f_0, t_0, \phi)$.
2. For each $\boldsymbol{\theta}^{(i)}$, determine the corresponding series of prevalence rates, $\boldsymbol{\rho}^{(i)} = M(\boldsymbol{\theta}^{(i)})$, by running the EPP model. This gives a sample from the induced prior $p^*(\boldsymbol{\rho})$ on the outputs.
3. Form the sampling importance weights for each $\boldsymbol{\rho}^{(i)}$ [and thus for each $\boldsymbol{\theta}^{(i)}$] as the product of the likelihood and the direct prior of $\boldsymbol{\rho}^{(i)}$:

$$w_i = p(\boldsymbol{W}|\boldsymbol{\rho}^{(i)}) \prod_{t \in \Gamma} I_{V_t}(\rho_t^{(i)}).$$

4. Sample from the discrete distribution of $\{\boldsymbol{\theta}^{(1)}, \ldots, \boldsymbol{\theta}^{(n)}\}$ with probabilities proportional to $w_i$ to approximate the posterior distribution for the inputs, and do the same for the outputs.

3.2. *Priors on input parameters.* We specify a joint prior distribution for the four input parameters $\boldsymbol{\theta} = (r, f_0, t_0, \phi)$. Separate sources of information are used to specify each of these priors, and so we specify these four parameters to be independent *a priori*. We assume that the rate of infection $r$ can take any value between 0 and 15 with equal probability, $r \sim U[0, 15]$, meaning that the average number of people an infected person infects per year can range from zero to 15 times the fraction at risk. The start year of the epidemic $t_0$ has a uniform discrete distribution on $\{1970, 1981, \ldots, 1990\}$.

The fraction initially at risk $f_0$ can be any value between zero and one, $f_0 \sim U[0,1]$. We assume a uniform prior on $[0,1]$ for the fraction entering the at-risk population $f(\frac{X(t)}{N(t)}, f_0, \phi)$. These two priors together define a prior for the behavioral response $\phi$. In the model, the fraction entering the at-risk group $Z(t)$ is given by

$$f\left(\frac{X(t)}{N(t)}, f_0, \phi\right) = \frac{\exp((\phi\chi(t)))}{\exp((\phi\chi(t))) - 1 + 1/f_0}, \tag{4}$$

where $\chi(t)$ is the difference between the fraction in the not-at-risk group at time $t$, namely, $X(t)/N(t)$, and the fraction $(1 - f_0)$ that was not at risk at the beginning of the epidemic. Thus,

$$\chi(t) = \frac{X(t)}{N(t)} - (1 - f_0). \tag{5}$$

Using (4), $\phi$ can be written in terms of the fraction entering the at-risk group and the fraction initially at risk. With uniform priors on both these fractions, the behavioral response parameter has a logistic distribution with mean $\frac{1}{\chi}\log(\frac{1}{f_0} - 1)$ and variance $\frac{\pi^2}{3\chi^2}$, and with an initial estimate $f_0 = 0.5$, it is centered around zero. The smaller the prior estimate of $\chi$ is, the more spread out the prior distribution for the behavioral response will be. We set $\chi = 0.1$ meaning that *a priori* we expect an average difference of 10% between the current and initial fractions not at risk.

3.3. *Likelihood for population prevalence.* In many of the countries with generalized HIV/AIDS epidemics, the data available for calibrating the EPP model consist of estimated prevalences among pregnant women at antenatal clinics. We will consider calibration of the EPP model in a country, or a part of a country, that is considered to be relatively homogeneous in terms of the pattern of the epidemic. UNAIDS often considers the urban areas of a country to form one such homogeneous part, and the rural areas to form another part. Here we assume that the prevalence among the attendees at an antenatal clinic gives an unbiased estimate of population prevalence (but see discussion in Section 5).

We now derive an output likelihood for the data, namely, the antenatal clinic prevalences, given the model outputs, namely, the population prevalences for each year during the observation period, $\boldsymbol{\rho} = (\rho_1, \ldots, \rho_T)$, where $\rho_t$ is the overall population prevalence in year $t$. The data consist of the number of infected women, $Y_{st}$, and the number of women tested, $N_{st}$, for clinic $s$ in year $t$, for the $S$ clinics $s = 1, \ldots, S$. Data are available for clinic $s$ for the years $t = t(s,1), \ldots, t(s, T_s)$, where $T_s$ is the number of years in which data were collected at clinic $s$.

We denote by $\gamma_{st}$ the prevalence at clinic $s$ in year $t$. We assume that $Y_{st} \sim \text{Binomial}(N_{st}, \gamma_{st})$, and that the $Y_{st}$ are conditionally independent of each other given the $\gamma_{st}$. Antenatal clinic data often include repeated measurements at the same clinic. To account for this repeated measurement structure, we approximate the likelihood by modeling $Y_{st}$ on the probit scale and using a hierarchical normal linear model. We let $W_{st} = \Phi^{-1}(x_{st})$, where $\Phi(\cdot)$ is the standard normal cumulative distribution function, and $x_{st} = (Y_{st} + \frac{1}{2})/(N_{st} + 1)$. The constants $\frac{1}{2}$ and 1 in the definition of $x_{st}$ are introduced for computational reasons, to avoid problems with zeros. They also have an approximate Bayesian motivation because $x_{st}$ would be the posterior mean of $\gamma_{st}$ with a noninformative Jeffreys Beta$(\frac{1}{2}, \frac{1}{2})$ prior distribution, in the absence of any other information.

Our model is then

$$(6) \qquad W_{st} = \Phi^{-1}(\rho_t) + b_s + \varepsilon_{st},$$

$$(7) \qquad b_s \stackrel{\text{iid}}{\sim} N(0, \sigma^2),$$

$$(8) \qquad \varepsilon_{st} \stackrel{\text{ind}}{\sim} N(0, v_{st}),$$

with

$$(9) \qquad v_{st} = 2\pi \exp\{\Phi^{-1}(\gamma_{st})^2\}\gamma_{st}(1 - \gamma_{st})/N_{st}.$$

In (6), $b_s$ is the clinic random effect for clinic $s$, assumed to be constant over time. The error term $\varepsilon_{st}$ approximates the binomial variation. Equation (9) is an approximation derived from the binomial distribution of $Y_{st}$ using the delta method.

Bayesian estimation of the model (6)–(8) requires a prior only for $\sigma^2$, and we use a standard Inverse Gamma ($\beta_1$, $\beta_2$) prior. To assess values of $\beta_1$ and $\beta_2$, we considered the results of fitting the model to urban antenatal clinic data from nine countries in sub-Sarahan Africa, and chose values of $\beta_1$ and $\beta_2$ that made this prior distribution spread out enough to amply cover the results of the model fits, namely, $\beta_1 = 0.58$, $\beta_2 = 93$.

We now use the model (6)–(8) to derive the likelihood of the data $\boldsymbol{W} = (\boldsymbol{W}_s : s = 1, \ldots, S)$, where $\boldsymbol{W}_s = (W_{st} : t = t(s, 1), \ldots, t(s, T_s))$, given the population prevalences $\boldsymbol{\rho} = (\rho_1, \ldots, \rho_T)$. This likelihood follows by integrating out the clinic random effects, $\boldsymbol{b} = (b_1, \ldots, b_S)$, and the random effects variance, $\sigma^2$, as follows:

$$(10) \quad \begin{aligned} p(\boldsymbol{W}|\boldsymbol{\rho}) &= \int \int p(\boldsymbol{W}|\boldsymbol{\rho}, \boldsymbol{b}) p(\boldsymbol{b}|\sigma^2) \, d\boldsymbol{b} p(\sigma^2) \, d\sigma^2 \\ &= \int \int \left\{ \prod_{s=1}^{S} p(W_s|\boldsymbol{\rho}, b_s) p(b_s|\sigma^2) \right\} d\boldsymbol{b} p(\sigma^2) \, d\sigma^2 \\ &= \int \left\{ \prod_{s=1}^{S} A_s(\sigma^2) \right\} p(\sigma^2) \, d\sigma^2, \end{aligned}$$

where $A_s(\sigma^2) = \int \prod_{s=1}^{S} p(W_s|\boldsymbol{\rho}, b_s) p(b_s|\sigma^2) \, db_s$.

The quantity $A_s(\sigma^2)$ can be evaluated analytically as follows. Let $d_{st} = W_{st} - \Phi^{-1}(\rho_t)$ and $\boldsymbol{d}_s = (d_{s,t(s,1)}, \ldots, d_{s,t(s,T_s)})$. Then $\boldsymbol{d}_s = b_s \boldsymbol{1}_{T_s} + \boldsymbol{\varepsilon}_s$, where $\boldsymbol{1}_{T_s}$ is a $T_s$-vector of ones and $\boldsymbol{\varepsilon}_s = (\varepsilon_{s,t(s,1)}, \ldots, \varepsilon_{s,t(s,T_s)})$. Thus, $\boldsymbol{d}_s \sim \text{MVN}_{T_s}(\boldsymbol{0}, \boldsymbol{\Sigma}_s)$, where $\boldsymbol{\Sigma}_s = \sigma^2 \boldsymbol{J}_{T_s} + \boldsymbol{V}_s$, with $\boldsymbol{J}_{T_s}$ defined as a $T_s \times T_s$ matrix all of whose elements are one, and $\boldsymbol{V}_s = \text{diag}(v_{st} : t = t(s,1), \ldots, t(s,T_s))$. It follows that $A_s(\sigma^2)$ is equal to the $\text{MVN}_{T_s}(\boldsymbol{0}, \boldsymbol{\Sigma}_s)$ density evaluated at $\boldsymbol{d}_s$. The error variance $v_{st}$ is approximated as in (9) with $\gamma_{st} = x_{st}$.

We have now reduced the integral (10) to a one-dimensional integral. This has no analytic solution and we evaluate it using numerical quadrature, specifically the globally adaptive interval subdivision method [Piessens et al. (1983)], as implemented in the R function `integrate`.

3.4. *Assessing model fit: Predictive distribution for clinical prevalence.* We assess the fit of the overall prediction procedure, including the EPP model itself, the Bayesian melding estimation method and the approximations we have used in deriving the likelihood, using out-of-sample predictive distributions. We are therefore interested in the predictive distribution of observed prevalence at clinic $s$ at a future time $u > T$. Clinic-specific prediction intervals can be used to evaluate the predictive quality of the EPP model by forecasting prevalence using data before a given point in time and comparing the predictive distributions with observed clinic prevalence for the years after that point, viewed as a test sample.

A sample from the posterior distribution of population prevalence for observed and future years, $p(\boldsymbol{\rho}|\boldsymbol{W})$, is given by the Bayesian melding output, $\boldsymbol{\rho}^{(1)}, \ldots, \boldsymbol{\rho}^{(J)}$, where $\boldsymbol{\rho}$ now includes future prevalences up to year $u$, so that $\boldsymbol{\rho} = (\rho_1, \ldots, \rho_T, \rho_{T+1}, \ldots, \rho_u)$. The predictive distribution of the future transformed observed prevalence at clinic $s$ and time $u$ is

$$p(W_{su}|\boldsymbol{W}) = \int \int p(W_{su}|b_s, \boldsymbol{\rho}) p(b_s|\boldsymbol{\rho}, \boldsymbol{W}) \, db_s p(\boldsymbol{\rho}|\boldsymbol{W}) \, d\boldsymbol{\rho}.$$

A Monte Carlo approximation to this is

$$\frac{1}{J} \sum_{j=1}^{J} p(W_{su}|b_s^{(j)}, \boldsymbol{\rho}_u^{(j)}),$$

where $b_s^{(j)}$ is one value sampled from $p(b_s|\boldsymbol{\rho}^{(j)}, \boldsymbol{W})$. This density is proportional to

$$(11) \quad p(b_s|\boldsymbol{\rho}^{(j)}, \boldsymbol{W}) = p(b_s|\boldsymbol{d}_s^{(j)})$$

$$(12) \qquad\qquad\qquad \propto p(\boldsymbol{d}_s^{(j)}|b_s) p(b_s)$$

$$(13) \qquad\qquad\qquad \propto \exp\left(-\frac{1}{2} \sum_{t=1}^{T_s} \frac{(d_{st}^{(j)} - b_s)^2}{v_{st}}\right) \left(\frac{1}{2} b_s^2 + \frac{1}{\beta_2}\right)^{-\beta_1 - 1/2},$$

because $\boldsymbol{d}_s^{(j)}|b_s \sim \text{MVN}_{T_s}(b_s\boldsymbol{1}_{T_s}, \boldsymbol{V}_s)$ and the marginal prior distribution of $b_s$ is proportional to $(\frac{1}{2}b_s^2 + \frac{1}{\beta_2})^{-\beta_1-1/2}$.

We sample from $p(b_s|\boldsymbol{d}_s^{(j)})$ using rejection sampling. Then the predictive distribution of the transformed observed prevalence is

$$W_{su}|b_s^{(j)}, \rho_u^{(j)} \sim N(\omega_u^{(j)} + b_s^{(j)}, v_{su}^{(j)}),$$

where $\omega_u^{(j)} = \Phi^{-1}(\rho_u^{(j)})$, and $v_{su}^{(j)}$ is given by (9) with $\gamma_{su} = \Phi(\omega_u^{(j)} + b_s^{(j)})$. Transforming the prevalence back to its original scale gives a sample from the posterior predictive distribution of the future observed prevalence.

**4. Results.** Data on prevalence at urban and rural antenatal clinics are available from the Epidemiological Fact Sheets on HIV/AIDS and Sexually Transmitted Infections 2006 at
http://www.who.int/globalatlas/predefinedReports/EFS2006/index.asp. As an example, we will discuss urban prevalence in Uganda, from where antenatal clinic prevalence data are available through 2002. Prevalence was observed at five clinics in Kampala and most of the observations were at two missionary hospitals. Prevalence has been falling since the early 1990s, as can be seen from Figure 2, which plots observed prevalence against time.

We ran the Bayesian melding procedure as described in the previous section for the five urban clinics in Uganda. We sampled 200,000 combinations of the input parameters from their prior distribution as described in Section 3.2. We restricted prevalence in 1980 to be smaller than 10%, $\rho_{1980} \sim U(0, 10)$. After calculating the weights for each of those inputs, we resampled 3,000 trajectories of which 373 were unique. The trajectory with the largest likelihood was resampled 120 times. Figure 2 shows the posterior sample of epidemic curves. The decrease in prevalence is projected to continue until around 2015 when it levels off between 1% and 6%, with a posterior median of about 3%. Figure 3 shows histograms of the samples of posterior predictive prevalence for 2005 and 2010. Although there is considerable uncertainty about the prevalence in any given future year, the results clearly predict a continuing overall decline in prevalence.

Figure 4 displays the prior density and the histogram of a sample from the posterior distribution of each of the four input parameters. The posterior mean of the rate of infection $r$ is 4.6 (95% confidence interval [1.8, 12.7]), meaning that on average each year an infected person infects a number of people equal to between 2 and 13 times the at-risk fraction of the population. The posterior median of the start year $t_0$ is 1979 (95% confidence interval [1972, 1983]). The fraction initially at risk $f_0$ is centered around 0.32 (95% confidence interval [0.26, 0.39]). The behavioral response $\phi$ is negative with a mean of $-5.6$ (95% confidence interval $[-9.3, -3.0]$). This means that as the number of AIDS deaths grows, the proportion of 15 year olds drawn into the at-risk group declines.

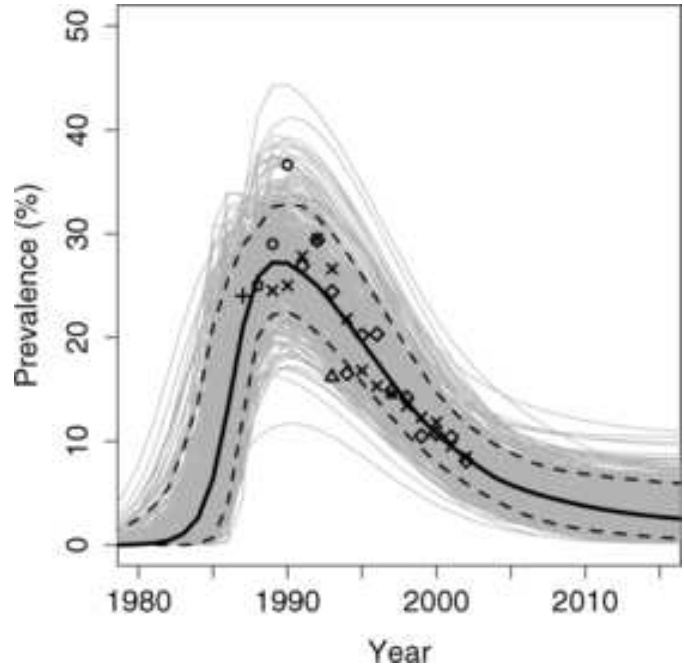

Fig. 2. *Posterior distribution of urban HIV prevalence in Uganda over time. Each dot is an observation, and dots with the same symbol correspond to repeated observations at the same clinic. Each grey line is a unique trajectory in the posterior sample of epidemic curves. The dashed lines are the 2.5% and 97.5% quantiles, and the solid black line is the median of the posterior sample.*

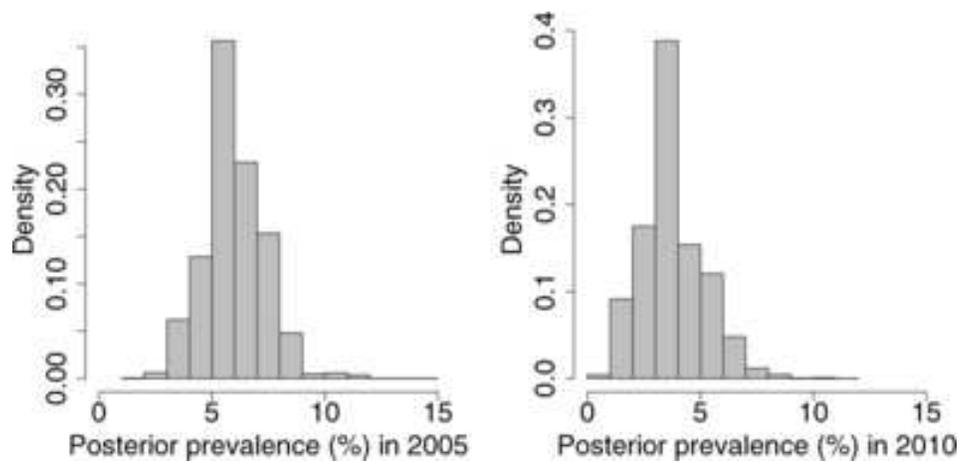

Fig. 3. *Sample from the posterior predictive distribution of urban HIV prevalence in Uganda in 2005 and 2010.*

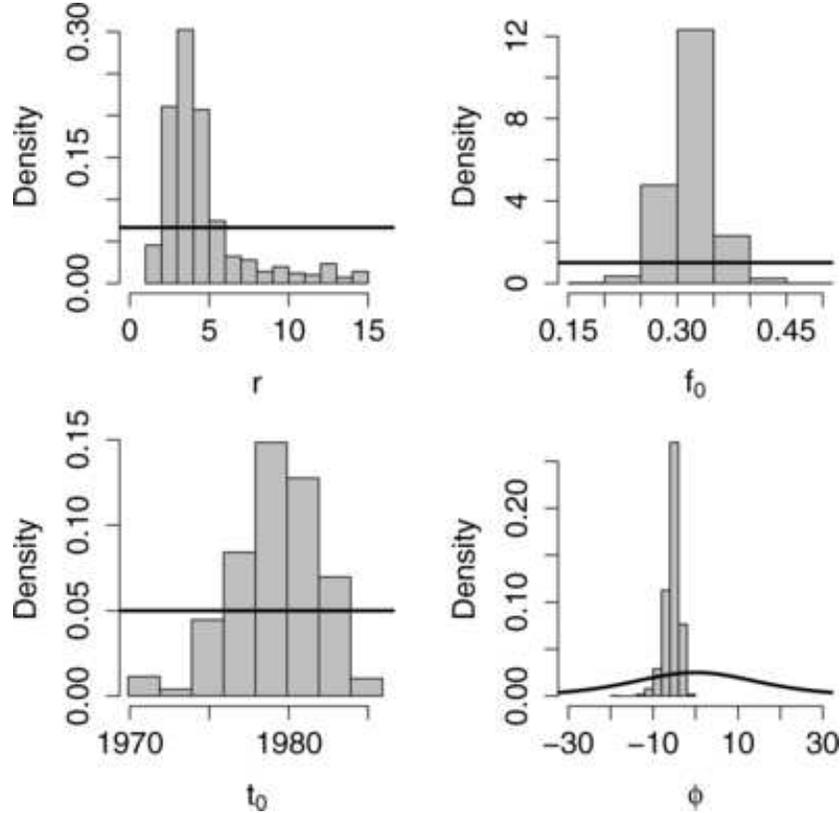

Fig. 4. *Posterior samples (histograms) and prior densities (solid lines or curve) of input parameters for urban prevalence in Uganda: (a) $r$; (b) $f_0$; (c) $t_0$; (d) $\phi$.*

Even though they are independent *a priori*, the four parameters of the EPP model are correlated *a posteriori*. Often the posterior relationship between the parameters is nonlinear. For example, in Figure 5 the rate of infection $r$ is plotted against the start year of the epidemic $t_0$. This shows the classic "banana shape" often seen in posterior distributions of the parameters of deterministic simulation models of this kind; see Raftery et al. (1995) for other examples. In this case the banana shape of the posterior arises because the data contain substantial information about the product of the rate of increase $r$ and the time since the start of the epidemic $(t - t_0)$ but less information on either $r$ or $(t - t_0)$ individually, as can be seen from Figure 4. As a result, the product $r(t - t_0)$ is fairly well identified, leading roughly to a reflected hyperbola in the plot of the joint posterior distribution of $r$ and $t_0$.

To assess the predictive performance of our method, we ran Bayesian melding using data from different truncated time periods, and compared the predicted to the observed prevalences for the remaining years, viewed as a test dataset. Figure 6 displays the prediction intervals for urban prevalence based on data through 1994, 1998 and 2002. Each grey line is a unique trajectory from the posterior sample of epidemic curves, the dotted lines are the 2.5% and 97.5% quantiles, and the solid black line is the median of the posterior sample for each year.

Figure 6(a) shows results based on observed prevalence through 1994 (the vertical line). At that time it was predicted that prevalence would decline, but there was a great deal of uncertainty about future prevalence. Based on the data through 1994 only, the 95% prediction interval for 2010 was [0%, 23%] with a median prediction of 1%. The subsequent clinic observations for 1995–2002 lay well within the posterior intervals for future overall prevalence. (In interpreting this result, it should be borne in mind that these bounds give predictive intervals for national urban prevalence and not for individual clinic observations.) Based on data through 1998, the posterior predictive distribution of prevalence in 2010 was much more concentrated, with a 95% interval [0%, 7%] instead of [0%, 23%], although the poste-rior predictive median was unchanged at 1%; see Figure 6(b). Again, the subsequent observations lay within the posterior bounds for future overall prevalence. Using data through 2002, the 95% prediction interval for 2010 is [2%, 7%], with median predicted prevalence of 4%; see Figure 6(c).

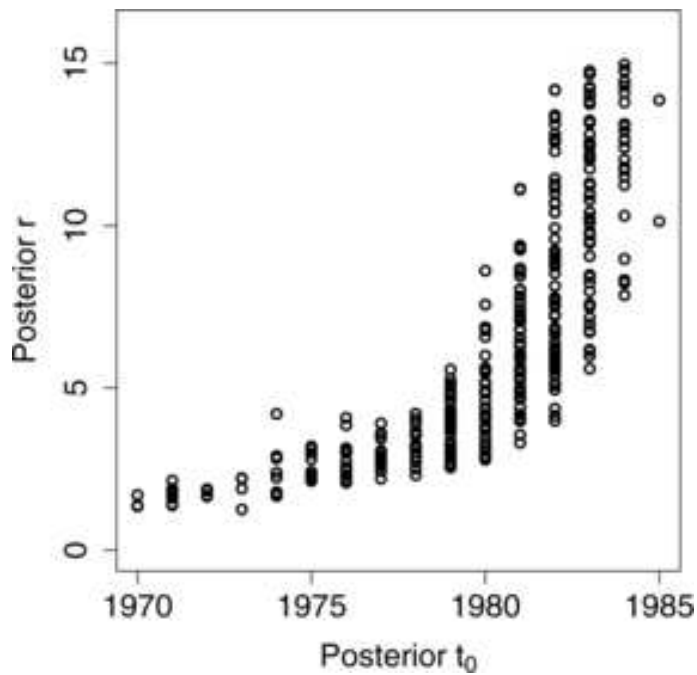

Fig. 5. *Sample from the joint posterior distribution of the rate of infection, $r$, and the start year, $t_0$, for urban prevalence in Uganda.*

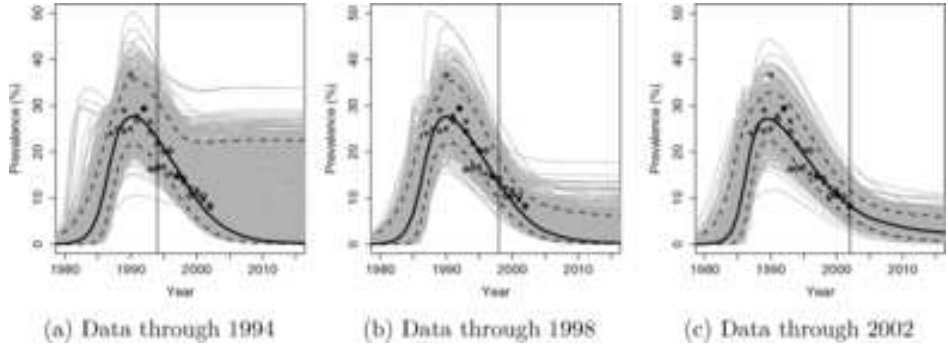

Fig. 6. *Posterior urban prevalence in Uganda, based on increasing observation periods.*

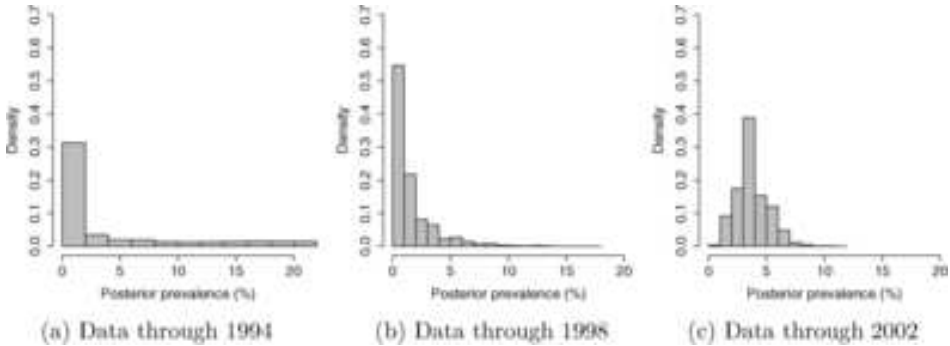

Fig. 7. *Histograms of samples from the posterior predictive distributions of urban HIV prevalence in Uganda in 2010, based on increasing observation periods.*

The 2010 posterior predictive distributions based on data through 1994, 1998 and 2002 are shown in Figure 7. As more data became available after 1998, prevalence was no longer predicted to decrease to zero.

We have used a prior on outputs that specifies prevalence in 1980 to be less than 10%. The result of the Bayesian melding procedure without this prior, using data up to 1994, is shown in Figure 8(b). This figure shows some prevalence curves in the posterior sample with high prevalence in the early 1980s. This contradicts expert knowledge about the HIV/AIDS epidemic, and illustrates the need for the direct prior distribution on outputs. This does not make much of a difference to the predictive distributions of future prevalence in individual years; for example, the prediction interval for 2010 is [0%, 24%] without the prior information on prevalence in 1980, compared to [0%, 23%] with the prior information; the predictive median is 1% for both. The main difference is that without the prior on outputs, the posterior includes unrealistic trajectories. Leaving out the prior on outputs does not change the posterior samples of prevalence curves based on data up to 1998 or 2002.

To assess the fit of our overall modeling procedure, including the EPP model itself, the Bayesian melding method and the approximations used in deriving the likelihood, we compared observed clinic prevalence with its predictive distribution, as described in Section 3.4. Figure 9 shows the predictive intervals for the two missionary hospitals in Kampala, Nsambya and Rubaga. Figures 9(a) and 9(d) display prediction intervals based on data through 1994, and Figures 9(b) and 9(e) display the results using data through 1998. For both hospitals at both time points, the observed future prevalences lay within their prediction intervals. These results are consistent with the statement that EPP model combined with the Bayesian melding procedure produces reasonable predictive intervals. Two trajectories of simulated clinic prevalence based on data through 1994 and 1998 are given in the same figures in grey, as well as in Figure 9(c) and 9(f) for data through 2002, showing that the modeling assumptions made in Section 3.3 represent the clinic prevalence data reasonably well.

**5. Discussion.** In this article we have proposed using Bayesian melding to obtain probabilistic projections of HIV prevalence from the EPP model developed by UNAIDS. EPP generates HIV prevalence trends for generalized epidemics and is typically fit to trends in the prevalence among women who attend antenatal clinics. For many countries in sub-Saharan Africa this is the only available information, a fact that precludes the use of more complicated models that require detailed information about the behavioral and biological determinants of an HIV epidemic.

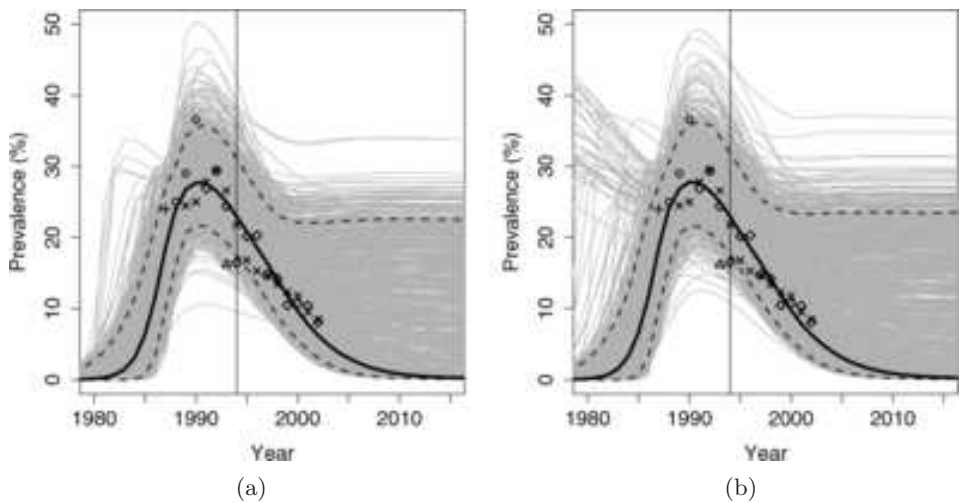

FIG. 8. *Posterior urban prevalence in Uganda, based on data up to 1994,* (a) *with a $U(0, 10\%)$ prior distribution on prevalence in 1980,* (b) *without prior information on prevalence.*

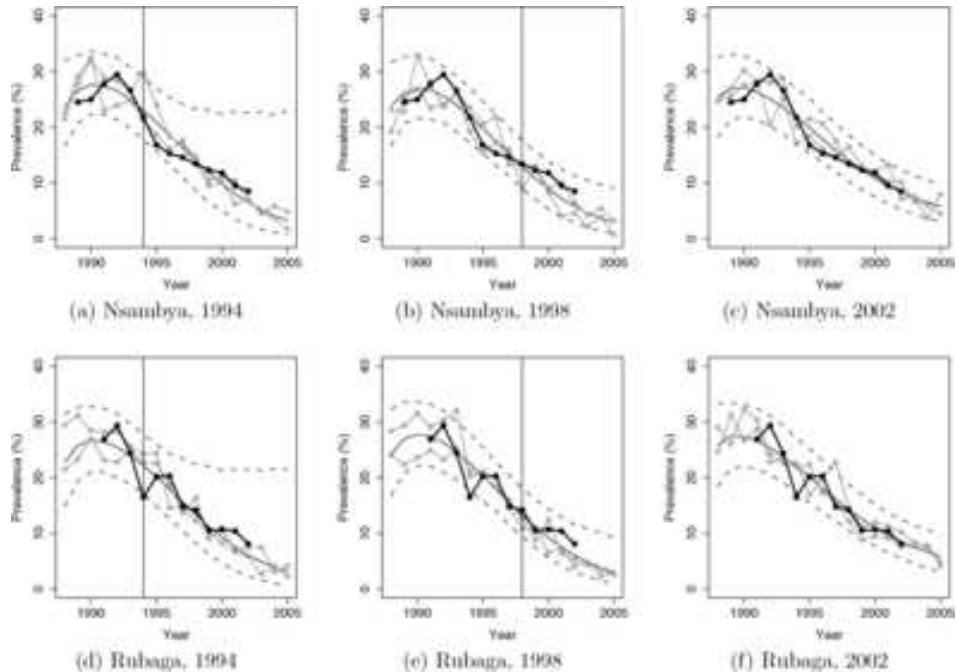

Fig. 9. *Predictive distributions and confidence intervals for the two urban clinics Nsambya and Rubaga in Uganda based on data through 1994, 1998 and 2002. The dotted lines are the 2.5% and 97.5% quantiles and the solid black line is the median of the predictive distribution for each year. The black line with the dots shows observed prevalence, and the grey lines show simulated observed clinic prevalence.*

Using the Bayesian melding procedure, uncertainty about future prevalence is described by building probabilistic projections. A random effects model is used to account for differences between clinics, and out-of-sample predictive distributions are derived and compared with data to assess the predictive quality of the overall modeling procedure, including the EPP model, Bayesian melding and the approximations involved in the likelihood.

We applied our method to Uganda where a relatively long series of antenatal clinic prevalence observations is available. The out-of-sample predictive performance of our method was good in this example, which lends support to the use of Bayesian melding for assessing uncertainty. This also shows that the relatively simple EPP model itself is effective at predicting prevalence, which is reassuring given how widely EPP is used. Further, it indicates that the approximations we used in deriving the likelihood have not unduly affected the procedure's predictive performance. These include our use of a normal distribution on the probit scale to approximate the binomial distribution. It would be possible to build a model that instead uses the binomial response explicitly, but this would be considerably more complicated, and our results suggest that the gain in performance from doing so would be modest at best.

We based our prevalence predictions on data from antenatal clinics, which is the usual way in which the EPP model is used. However, it has been convincingly argued that prevalence estimates from antenatal clinic data tend to be biased upward. There are several reasons for this. Young women, making up a substantial fraction of all pregnant women, tend to have higher HIV prevalence than the general population [Zaba et al. (2000)]. Also, urban antenatal clinics tend to be public rather than private and, as a result, oversample poorer women who are more likely to be HIV positive. Further, rural antenatal clinics often underrepresent remote rural areas that tend to have lower prevalence and where a large fraction of the population typically lives [Saphonn et al. (2002), Boerma et al. (2003)].

Nationally representative population surveys have been proposed as an alternative to antenatal clinic data, particularly the Demographic and Health Surveys (DHS) [Boerma et al. (2003)]. For countries with national HIV prevalence estimates, a bias term was included in the 2005 version of EPP to calibrate the estimates that it produces. There are problems with this also. The DHS estimates themselves tend to be biased downward largely due to nonresponse. People not living in households, who are often more likely to be HIV positive, including sex workers living in brothels, are underrepresented because the DHS is a household survey [Lydie et al. (2004), Zaba et al. (2004), Mishra et al. (2006)]. The DHS estimates can also be highly variable, particularly in low prevalence countries where they may be based on a relatively small number of HIV positive cases.

Future work will extend Bayesian melding for EPP to take account of the DHS prevalence estimates and information describing possible bias in both antenatal clinic and DHS data. One possible way of doing this would be to include a second likelihood based on DHS data, as well as bias terms for both antenatal clinic and DHS data. Priors for the bias terms could be set using data from other countries.

More generally, we have used data from a range of countries to set the priors used. We have done this rather informally, but have got good results in terms of out-of-sample prediction. A more formal and comprehensive way of doing this would be to build a Bayesian hierarchical model for the data on a set of comparable countries, with country being an additional level of the hierarchy. We did not use such a model here because it is important that officials from each country be able to develop and modify their own estimates and projections of HIV/AIDS. Thus, UNAIDS provides tools for them to do this, along with suggested defaults, including suggested priors. In practice, this means that a one-country approach must be used. A more comprehensive multi-country Bayesian hierarchical model would be a reasonable project for future work and could help set the priors for individual countries in a more formal and reproducible way. However, our results are relatively insensitive to changes in the priors we have used, and so it seems unlikely that a more comprehensive approach would give very different results from the one-country method we have developed here. The obstacles to a more comprehensive multi-country modeling approach may be more political than statistical.

When constructing HIV projections in the future, the availability and possible effects of antiretroviral therapy (ART) will also need to be taken into account. So far the availability of ART is limited and the effects negligible. Recently, Botswana and South Africa were two of the first countries in sub-Saharan Africa to establish national ART programs. Other possible model improvements for EPP include additional parameters to model behavioral change, such as time variation in the force of infection to deal with rapid changes in behavior occurring in some countries [Ghys et al. (2006)]. However, it is not clear that making the model more complicated will improve its performance; whether or not this is the case in the presence of limited data is an empirical question.

**Acknowledgments.** The authors are grateful to Jeff Eaton for research assistance, to Tim Brown, Peter Ghys, Nick Grassly, Eleanor Gouws, Meade Morgan, Josh Salomon and Karen Stanecki for helpful discussions, and to the Associate Editor for useful comments that improved the paper. Alkema and Raftery thank ÚTIA, the Institute of Information Theory and Automation, Prague, Czech Republic, for hospitality.

L. Alkema
A. E. Raftery
Department of Statistics
University of Washington
Seattle, Washington 98195-4322
USA
E-mail: alkema@u.washington.edu
        raftery@u.washington.edu

S. J. Clark
Department of Sociology
University of Washington
Seattle, Washington 98195-3340
USA
E-mail: samclark@u.washington.edu